\documentclass[aip,apl, amsmath,amssymb, reprint,]{revtex4-1}%
\usepackage{graphicx}% Include figure files
\usepackage{dcolumn}% Align table columns on decimal point

\begin{document}%

\title{Reconfigurable vortex beam generator based on the Fourier transformation principle}
\author{Aiping Liu}
\affiliation{Affiliation: Institute of quantum information and technology, Nanjing University of Posts and Telecommunications, Nanjing 210003, China, and Key Lab of Broadband Wireless Communication and Sensor Network Technology, Nanjing University of Posts and Telecommunications, Ministry of Education, Nanjing 210003, China}
\author{Chang-Ling Zou}
\email{clzou321@ustc.edu.cn}
\author{Xifeng Ren}
\affiliation{Affiliation: Key Laboratory of Quantum Information, University of Science and Technology of China, CAS, Hefei, 230026, China, and Synergetic Innovation Center of Quantum Information and Quantum Physics, University of Science and Technology of China, Hefei, 230026, China}
\author{Wen He}
\author{Mengze Wu}
\affiliation{Affiliation: Institute of quantum information and technology, Nanjing University of Posts and Telecommunications, Nanjing 210003, China, and Key Lab of Broadband Wireless Communication and Sensor Network Technology, Nanjing University of Posts and Telecommunications, Ministry of Education, Nanjing 210003, China}
\author{Guangcan Guo}
\affiliation{Affiliation: Key Laboratory of Quantum Information, University of Science and Technology of China, CAS, Hefei, 230026, China, and Synergetic Innovation Center of Quantum Information and Quantum Physics, University of Science and Technology of China, Hefei, 230026, China}
\author{Qin Wang}
\affiliation{Affiliation: Institute of quantum information and technology, Nanjing University of Posts and Telecommunications, Nanjing 210003, China, and Key Lab of Broadband Wireless Communication and Sensor Network Technology, Nanjing University of Posts and Telecommunications, Ministry of Education, Nanjing 210003, China}

\begin{abstract}
 A method to generate the optical vortex beam with arbitrary superposition of different orders of orbital angular momentum (OAM) on a photonic chip is proposed. The distributed Fourier holographic gratings are proposed to convert the propagating wave in waveguides to a vortex beam in the free space, and the components of different OAMs can be controlled by the amplitude and phases of on-chip incident light based on the principle of Fourier transformation. As an example, we studied a typical device composed of nine Fourier holographic gratings on fan-shaped waveguides. By scalar diffraction calculation, the OAM of the optical beam from the reconfigurable vortex beam generator can be controlled on-demand from $-2_{nd}$ to $2_{nd}$ by adjusting the phase of input light fields, which is demonstrated numerically with the fidelity of generated optical vortex beam above 0.69. The working bandwidth of the Fourier holographic grating is about 80 nm with a fidelity above 0.6. Our work provides an feasible method to manipulate the vortex beam or detect arbitrary superposition of OAMs, which can be used in integrated photonics structures for optical trapping, signal processing, and imaging.
%\keywords{optical angular momentum, holographic grating, vortex beam; OCIS codes: (090.2890) Holographic optical elements; (130.0130) Integrated optics; (230.7370) Waveguides}
\end{abstract}
%\newpage
\maketitle

\section{INTRODUCTION}

Orbital angular momentum (OAM) of a single photon has inherent infinite and orthogonal dimension for encoding quantum information, which is proposed to be a fascinating area of research since 1992. \cite{1Allen1} Because of the unbounded dimensions of the OAM, it is possible to encode a single photon by OAM in a high dimensional space. \cite{2Mair2,3Ren3,4Yu4} The potential application of OAM in quantum attract a lot of interest. Efficient OAM generation, \cite{5Dwyer5,6Heffernan6} OAM entanglement storage, \cite{7Ding7} OAM data transmission \cite{8Wang8} and OAM detection \cite{9Liu9} have been realized recently.

As the real-time manipulation of the OAM components of a vortex beam is necessary for various applications, \cite{10Ionicioiu10,11Wang11} it is desirable for a re-configurable vortex beam generator. Among various experiment platforms for vortex beam generation, the photonic chip approach is of great interest because of its high stability and scalability. \cite{12Cai12,13Liu13} In order to be compatible with present photonic integrated circuit fabrication technology, methods of generating OAM based on the light scattering by gratings in the cavities can be converted to free space beam with non-zero OAM. \cite{4Yu4,12Cai12} Alternatively, the holographic method has been proposed, however, the order of OAM obtained by holography is determined by the geometry of the holographic grating, and is not reconfigurable when for a given holography. \cite{14He14} It is proposed that by putting two holography grating on two separate waveguides, the generated beam profile can be controlled by adjusting the input lights in two waveguides, which points out a promising approach for reconfigurable vortex beam generation. \cite{12Cai12} Recently, the waveguide-based holography grating has been demonstrated, \cite{15Nadovich15,16Zhou16} representing the first step towards reconfigurable vortex beam generation and detection.

In this work, a reconfigurable vortex beam generator (RVBG) on a photonic chip is proposed, which is based on the Fourier holographic gratings connected to an array of waveguides. Arbitrary superposition of OAM in an optical vortex beam can be generated by controlling the phase and amplitude of lights in the waveguides, or be detected as a reversal process. As an example, we show the procedure to construct the RVBG by nine Fourier holographic gratings on nine fan-shaped waveguides ranged in a disk. By numerical simulation, we demonstrated that the vortex beam with OAM from $-2_{nd}$ to $2_{nd}$ can be selectively generated with high fidelity and large wavelength bandwidth. Besides of the vortex beam with single OAM, vortex beams with a superposition of different OAMs can also obtained by the proposed device with appropriated incident light phases and amplitudes in waveguides.

\section{PRINCIPLE}

As shown in Fig. 1(a), the proposed RVBG is composed of nine fan-shaped waveguides ranged in a disk, where the holographic gratings are fabricated on the top of the waveguides near the center of the disk. In this RVBG, the propagating waves in the fan-shaped waveguides (in plane) will be scattered by the holographic grating, and the scattered light form the vortex beam in free space (in vertical direction) together. What is more, the generated beam is reconfigurable that the components of different OAMs can be manipulated by controlling the amplitude and phase of the input lights, because the vortex beam generated by the holographic grating is based on the principle of Fourier transformation.

The basic principle of the Fourier holographic grating can be explained as following: the holographic gratings on each fan-shaped waveguide can generate a superposition of OAM beams. For $(2N+1)$ -fan-shaped waveguide components, based on the Fourier principle, the $j_{th} (j\in[0,2N])$ waveguide can generate a beam as a superposition of OAMs

\begin{equation}
\label{e1}
E_{b,j}(r,\phi)=A_{j}\sum_{l=-N}^{N}F(l,r)e^{il\phi}e^{ij(l+N)\frac{2\pi}{2N+1}},
\end{equation}
where $(r,\phi)$ are the cylindrical coordinator,  $A_{j}$ is the amplitude of the input light in $j_{th}$ waveguide, $F(l,r)$ is the radial field distribution of vortex beam with OAM of $l_{th}$ order $(l\in[-N,N])$. Therefore, when the input lights to the waveguides are with equal intensity but with a phase gradient of $n\Delta (\Delta=\frac{2\pi}{2N+1})$, i.e.
\begin{equation}
\label{e2}
A_{j}=Ae^{-ijn\Delta},
\end{equation}
then the generated beam should be
\begin{equation}\label{e3}
E(r,\phi)=\sum_{j}E_{b,j}(r,\phi)=A\cdot F(n,r)e^{in\phi}.
\end{equation}
It means that only the $n_{th}$ OAM can be generated as the constructive interference between the scattered light from all gratings, while the component of other OAM are suppressed due to the destructive interference. In reversal, if $l_{th}$ OAM beam shine on the fans, then the input light can be converted to all waveguide modes, but with a phase gradient of $l\Delta$.

By controlling the amplitude and phases of the input light, we can generate the arbitrary superposition of the OAM beams. If the input photon in the superposition of different paths be represented by a vector $\left|\Phi\right\rangle _{in}=\left\{ A_{j}\right\} ^{T}$, the output of RVBG is
\begin{equation}
\label{e4}
\left|\Phi\right\rangle _{out}=U_{F}\left|\Phi\right\rangle _{in},
\end{equation}
with $U_{F}$ is the unitary matrix for Fourier transformation, $\left|\Phi\right\rangle _{out}$ is defined on the basis of OAMs.

If we prepare a multimode interferometer (MMI) \cite{17Brooks17} for inverse Fourier transformation on the same chip and guide all the output of the waveguides to the MMI as shown in Fig. 1(b), then
\begin{equation}
\label{e5}
\left|\Phi\right\rangle _{out}=U_{F}U_{F}^{+}\left|\Phi\right\rangle _{in}=\left|\Phi\right\rangle _{in},
\end{equation}
we can utilize the Fourier transformation principle again to convert the superposition of OAMs inputs to superposition of light outputs in different paths, i.e. the input free-space photon with $l_{th}$ OAM be converted to light output in $l_{th}$ port. In reverse, the input light from $l_{th}$ waveguide can be converted to vortex beam of $l_{th}$ OAM. Generally, the photonic integrated circuits allows the real-time control of the phase, combining with the arbitrary unitary evolution in linear optics, reconfigurable waveguide mode converter can be realized. \cite{18Reck18,19Carolan19,20Flamini20,21Clements21} Therefore, arbitrary conversion between the waveguide paths to the OAMs can be realized if the reconfigurable mode converter $U_{R}=U_{F}^{\dagger}U$ with $U$ be the demanded arbitrary unitary, i.e.
\begin{equation}
\label{e6}
\left|\Phi\right\rangle _{out}=U_{F}U_{R}\left|\Phi\right\rangle _{in}=U\left|\Phi\right\rangle _{in}.
\end{equation}

Reversely, we have $\left|\Phi\right\rangle _{in}=U^{\dagger}\left|\Phi\right\rangle _{out}$, the vortex beam input to the RVBG can be converted to the arbitrary superposition of output in the waveguide, could be used for detecting and analyzing the OAM components of the vortex beam.

Here, we also provide the detailed procedure to calculate the fan-shaped waveguide grating. For the waveguide $h_{j}(j=0,1...8)$, a combination of vortex beam is incident vertically to the waveguide $h_{j}$ on the center of the round, while the other eight waveguides are shielded from the target vortex beam by an opaque as shown in Fig. 1(c). At the same time, a guiding wave propagates in the waveguide from outside to inside of the fan, which will interfere with the combination of vortex beam. The guiding wave in $h_{j}$ can be given as $A_{0j}=A^{\ast}e^{-ik_{j}r}$, where $k_{j}$ is the wave vector. The combination of vortex beam to form the holographic grating on the $j_{th}$ waveguide is a superposition of the target optical vortex beams with different orders and phase, which has the electric field form as $E_{b,j}(r,\phi)=A\sum_{l=-4}^{4}F(l,r)e^{il\phi}\times e^{ijl\Delta}$. For the RVBG, the phase difference of different vortex components changes according to the Fourier transformation, which guarantees that the output of nine holographic gratings are orthogonal to each other. The interference of the waves from two directions lead to string with black and bright patterns on the surface of the waveguides. The interference pattern on each waveguide is formed separately. The Fourier holographic grating is generated by dividing the interference region into many pixels $(60nm\times60nm)$ and setting a gray value of $G(x,y)$ for each pixel. \cite{22Liu22,23Chen23} According to the phase difference $\delta\theta$ between the target vortex beam and the guiding wave, the simple binary function is applied, i.e., $G(x,y)=0$ if $\delta\theta<0.5\pi$, or else $G(x,y)=1$. The inset of Fig. 1(c) gives the binary gray image of the Fourier holographic grating.

\begin{figure}\label{fig_1}
\centering\includegraphics[width=1\columnwidth]{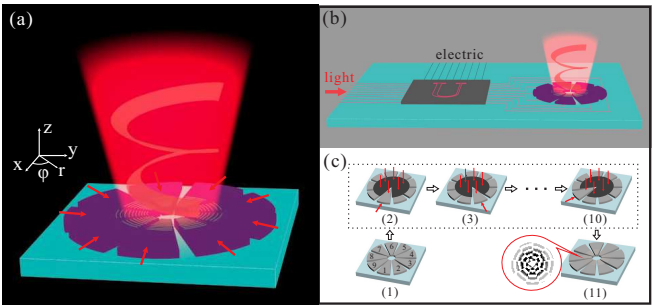} \caption{(color online). The schematic illustration of the Fourier holographic grating. (a) The vortex beam generated by nine incident wave couple to the fan-shaped holographic gratings on a chip. (b) The generation of vortex beam with an arbitrary superposition of OAMs by combining holographic gratings and arbitrary unitary mode converters. (c) The precedure for preparating the Fourier holographic gratings. The combination of target vortex beam and the guiding wave interfere to form the Fourier holographic grating on the $h_{j}$ waveguide with the other waveguides shielded from the combination of target vortex beam. The inset of (11) is the binary gray image of the Fourier holographic grating.}
\end{figure}

\section{RESULTS}

To verify the proposed Fourier holographic grating scheme, we performed numerical simulation based on a practical experimental model. The Fourier holographic grating is composed of $\textrm{Si}_{3}\textrm{N}_{4}$ waveguide on a silica substrate, with the refractive index of $\textrm{Si}_{3}\textrm{N}_{4}$ and silica as 2.035 and 1.45 at the wavelength of 670 nm. \cite{24Luke24} The field spatial distribution of generated beam is numerically calculated by the scalar diffraction theory.

In the following, we do not consider the waveguide mode converter, just simply assume we can have arbitrary input to the waveguide. Due to the principle of holography, when the RVBG is constructed by nine gratings $\left(N=4,\,\Delta=2\pi/9\right)$, nine guiding waves propagate along the waveguides together to produce the target vortex beam. To interact with the Fourier holographic grating, the form of the electric field propagating in the waveguide is set to be $A_{j}e^{-i\theta_{j}}$, where $\theta_{j}$ is the phase light incident on the guiding wave. According to the principle of holography, the electric field of the generated light is $E_{b}(r,\phi)=\sum_{j=0}^{8}E_{b,j}(r,\phi)$, which can be estimated as
\begin{equation}
\label{e7}
E_{b}(r,\phi)\propto\sum_{l=-4}^{4}F(l,r)e^{il\phi}(\sum_{j=0}^{8}A_{j}e^{ij(l+4)\Delta}),
\end{equation}
leads to a superposition of OAMs. The scattered lights together constitute the vortex beam with appropriate $\theta_{j}$, as shown in Fig. 2, in which the result is obtained on the plane with a distance of $10\lambda$ from the up surface of the waveguide. The waist of the target vortex beam is posited on the top surface of the waveguide with a diameter of 1 $\mu$m. When $\theta_{j}=2\Delta$, an optical vortex beam with $-2_{th}$ OAM is obtained, which has a donut-shaped amplitude distribution [Fig. 2(a2)] and a helical phase of $-4\pi$ with a singular in the center [Fig. 2(a4)]. When $\theta_{j}$ is changed to be $\theta_{j}=3\Delta$, the order $l$ of the optical vortex beam turns into -1, in which there is amplitude distribution deviated from the donut-shape [Fig. 2(b2)] and a helical phase of $-2\pi$ for the phase distribution [Fig. 2(b4)]. A Gauss beam is obtained with $\theta_{j}=4\Delta$ as shown in Fig.2(c2) and 2(c4), where the donut-shaped amplitude distribution and the helical phase disappear. When $\theta_{j}$ is set to be $5\Delta$, the generated optical vortex beam with $1_{st}$ OAM is obtained as shown in Fig.2(d2) and Fig.2(d4), in which there is an opposite helical phase of $2\pi$ compared to the case of $-1_{st}$ OAM. For $\theta_{j}=6\Delta$, an optical vortex beam with $2_{nd}$ OAM is generated, which has a big donut-shaped amplitude distribution and a helical phase of $4\pi$ as shown in Fig.2(e2) and 2(e4), respectively. The relation between $\theta_{j}$ and the order $l$ of the generated OAM is given in Tab.1. With the Fourier holographic grating, the order $l$ of the generated OAM can be manipulated freely by the phase difference $\theta_{j}$ of wave incident on the waveguides.

\begin{figure}\label{fig_2}
\centering\includegraphics[width=1\columnwidth]{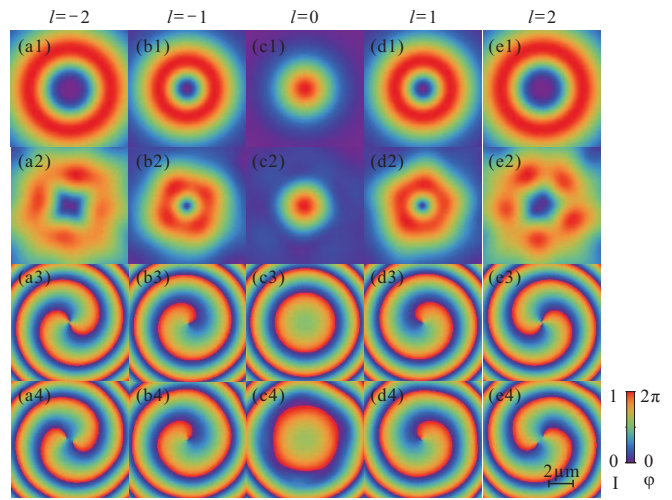} \caption{(color online). The amplitude and phase distributions of the optical vortex beams: (a1)-(e1) and (a2)-(e2) are the amplitude distributions for the target and generated optical vortex beams, respectively. (a3)-(e3) and (a4)-(e4) are the phase distributions for the target and generated optical vortex beams, respectively. The columns from left to right are the results of the optical vortex beams with $l=-2,-1,0,1,$ and $2$, respectively.}
\end{figure}

To qualify the quality of the generated vortex beam, the fidelity is introduced as \cite{22Liu22}
\begin{equation}\label{e8}
\mathcal{F}=\frac{|\int E^{\ast}(x)E_{t}(x)dx|^{2}}{\int |E(x)|^{2}dx^{3}\int |E_{t}(x)|^{2}dx^{3}},
\end{equation}

where $E(x)$ and $E_{\mathrm{t}}(x)$ are the amplitudes of the generated and target optical vortex beams, respectively. Because of the finite number of imaging pixels, the difference exists between the obtained beam and the target optical vortex beam, which makes the fidelity $\mathcal{F}<1$. Table 1 summarizes the fidelities of the generated OAM with different orders, in which a circular area with the electric field intensity attenuating to be $1/10$ of the maximum is selected to calculate the $\mathcal{F}$. When the order $l$ of OAM being manipulated between $-2$ to $2$, the fidelity of the generated vortex beam can be kept above 0.69. For $|l|=1$ and $2$, the generated vortex beam has relative high fidelities of about 0.82 and 0.91, respectively. While $|l|=0$, the generated vortex beam has a relatively low fidelity, which is attributed to the blank core of the fan-shaped holographic grating.

\begin{table}[!htb]
\renewcommand{\arraystretch}{1.3}
\caption{The order and the fidelity of the generated OAM for different incident phase difference $\Delta\phi$.}\label{tab_1} \centering
\begin{tabular}{cccccc}
\hline  $\Delta\phi$ & $4\pi/9$ & $6\pi/9$ & $8\pi/9$ & $10\pi/9$ & $12\pi/9$ \\
\hline  $l$ & -2 & -1 & 0 & 1 & 2 \\
\hline  $F$ & 0.9127 & 0.8235 & 0.6967 & 0.8202 & 0.9082 \\
\hline
\end{tabular}
\end{table}

With the Fourier holographic grating, the OAM components of generated optical vortex beam can be reconfigured from $-2_{nd}$ to $2_{nd}$ order freely. Further, we studied the property of Fourier holographic grating working on different wavelengths. The Fourier holographic grating is obtained with the interference of two waves at $\lambda_{0}=670$ nm in free space, and the light incident on the waveguide to generate the optical vortex beam is changed. Figure 3(a) gives the fidelity of the generated OAM as a function of the incident wavelength for the five different orders. The highest fidelity is obtained near 670 nm as designed, and the fidelity reduces when the wavelength of the light incident on the waveguide is far away from 670 nm. From Fig. 3(a), it is shown that the Fourier holographic grating has a working bandwidth of 80 nm (from 640 nm to 720 nm) with the fidelities above 0.6 for all of the five orders. So the designed Fourier holographic grating is practical for the information processing tasks.

Besides of the vortex beam with single OAM order, the Fourier holographic grating also can generate the beam of arbitrary superposition of OAMs. When the light input on the waveguides is a superposition of waves with additional phase shift $\phi_{j}$ to the $j\Delta$, which can be represent as
\begin{equation}\label{e9}
A_{wj}^{'}=A_{0}^{\ast}e^{-ij\Delta-i\phi_{j}},
\end{equation}

an optical beam consists of a superposition of OAMs can be obtained. For example, when $\phi_{1}=8\pi/9$ and $\phi_{2}=10\pi/9$, a superposition of $0$ and $1_{st}$ OAM is obtained as shown in Fig. 3(b). Figure 3(c) is the corresponding target vortex beam with the intensity distribution on the left column and phase distribution on the right column. The fidelity of the generated vortex beam with a superposition of $0$ and $1_{st}$ OAM is about 0.5234. Setting $\phi_{1}=10\pi/9$ and $\phi_{2}=12\pi/9$, a superposition of $1_{st}$ and $2_{nd}$ OAM is obtained with a fidelity of 0.7792 as shown in Fig. 3(d). The superposition states of OAM with arbitrary orders can be generated by changing the wave phase incident on the waveguides.

The generation of vortex beam with arbitrary superposition of OAMs broadens the application of OAM, since quantum superposition plays a key role in quantum communication. The OAM also provide larger Hilbert space for encoding information by a single photon, could boost the rate of key generation. Combine with the advantages of the photonic integrated circuits, the ultra-fast reconfigurability and scalability allow the stable and compact photonic chip for high-speed signal emitting, receiving and processing.

\begin{figure}\label{fig_3}
\centering
\includegraphics[width=1\columnwidth]{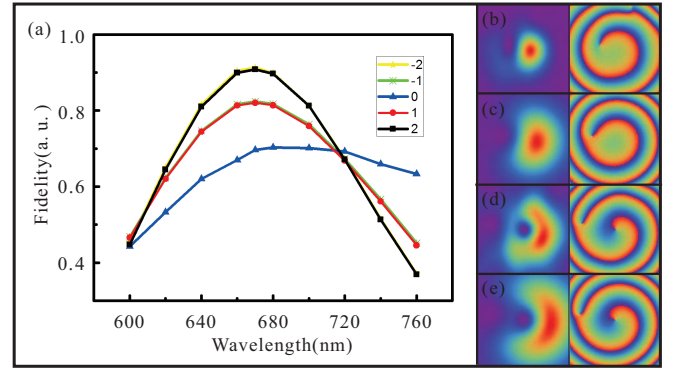}
\caption{(color online) (a) The fidelity of the generated vortex beam as a function of the working wavelength. Black (square), red (circular), blue (triangle), green (double cross) and yellow (star) lines are the fidelities of the vortex beam with the OAM be $l=-2,-1,0,1,$ and $2$, respectively.(b) and (c) are the generated and target vortex beam with a superposition of $l=0$ and $1$. (d) and (e) are the generated and target vortex beam with superposition of $1_{st}$ and $2_{nd}$ OAM. The left and right columns are the intensity and phase distributions of the electric field, respectively.}
\end{figure}

\section{CONCLUSION}
A Fourier holographic grating is proposed to generate the optical vortex beam with reconfigurable arbitrary superposition of orbit angular momentum. The switching between the OAM orders of $-2, -1, 0, 1,$ and $2$ is demonstrated numerically by controlling the phase of incident light based on the Fourier transformation principle. This Fourier holographic grating can work with a fidelity above 0.69 for all of the five orders and a working bandwidth of 80 nm. A reconfigurable vortex beam generator makes the information processing based on the OAM encoding more feasible for practical applications, since our device posses the advantages as reconfigurable, stable, compact and scalable.

\section{Acknowledgements}
This work was supported by the National Key R \& D program (Grant No.2016YFA0301300), the National Natural Science Foundation of China (Grant No. 11504183, 61505195, 61590932, 11774333, 61475197, 11774180), the Anhui Initiative in Quantum Information Technologies (No. AHY130300), the Scientific Research Foundation of Nanjing University of Posts and Telecommunications (NY214142), and the Fundamental Research Funds for the Central Universities.

\end{document}